%% file: main.tex
\documentclass[%
 reprint,
 superscriptaddress,
 amsmath,amssymb,
 aps,
]{revtex4-2}
\usepackage{array}
\usepackage{amsmath}
\usepackage{multirow}
\usepackage{tabularx, makecell, booktabs} 
\usepackage{graphicx}
\usepackage{dcolumn}
\usepackage{bm}
\usepackage{xcolor}
\newcolumntype{L}{>{$}l<{$}} 
\usepackage{hyperref}

\begin{document}
\preprint{APS/123-QED}
\title{Characterization of the Three-Flavor Composition of Cosmic Neutrinos with IceCube}


\input{authorlist}

\date{\today}

\collaboration{IceCube Collaboration}
\noaffiliation

\begin{abstract}

Neutrinos oscillate over cosmic distances. Using 11.4 years of IceCube data, the flavor composition of the all-sky neutrino flux from 5\,TeV--10\,PeV is studied. We report the first measurement down to the $\mathcal{O}$(TeV) scale using events classified into three flavor-dependent morphologies. The best fit flavor ratio is $f_e:f_{\mu}:f_{\tau}\,=\,0.30:0.37:0.33$, consistent with the standard three-flavor neutrino oscillation model. Each fraction is constrained to be $>0$ at $>$ 90\% confidence level, assuming a broken power law for cosmic neutrinos.
We infer the flavor composition of cosmic neutrinos at their sources, and find production via neutron decay lies outside the 99\% confidence interval.

\end{abstract}

\maketitle


%
%
\textbf{Introduction:} 
High energy astrophysical phenomena in the Universe are typically investigated via measurements of the energy spectra of photons, neutrinos, and cosmic rays. A unique way of probing the cosmic-ray production and acceleration mechanisms of astrophysical sources is via measurements of the astrophysical neutrino flavor composition, which determines the fractional contribution of electron, muon, and tau neutrinos, denoted as $f_e$, $f_{\mu}$ and $f_{\tau}$, respectively.
A measurement of the flavor ratio of cosmic neutrinos on Earth, accounting for neutrino oscillations, allows us to infer the flavor ratios at which cosmic neutrinos are produced in the dominant source populations, and constrain the properties of the production environments in these sources.
In case of neutrino production via pion decay (when all charged pions and muons decay, without significant energy loss or gain), we can expect a ratio of $f_{e,\mathrm{S}}:\,f_{\mu,\mathrm{S}}:\,f_{\tau,\mathrm{S}}\,=\left(1:2:0\right)_{\mathrm{S}}$ generated at the source, including both neutrinos and antineutrinos~\cite{Becker:2007sv}. However, these neutrinos undergo oscillations as they traverse the Universe and arrive at Earth, where we expect their composition to be $f_{e,\oplus\!}:\,f_{\mu,\oplus\!}:\,f_{\tau,\oplus\!}\, = \left(0.30:0.36:0.34\right)_{\oplus\!}$, assuming the mixing parameters from~\cite{Esteban:2020cvm}.
Any deviation from this scenario will result in a different observed flavor ratio.
For example, interactions including physics beyond the standard theory of oscillations or a violation of the principle of Lorentz invariance can cause deviations in the observed ratio~\citep{new_physics,Bustamante:2016ciw,Carloni:2022cqz,Rasmussen:2017ert}. 
The observed ratio will also differ from the standard expectation 
if the neutrino production mechanism at the source is inconsistent with the typical pion decay scenario. 
In case of a dominant neutron decay mechanism~\cite{Anchordoqui:2003vc,PhysRevD.75.123005}, the flavor ratio will be $\left(1: 0: 0\right)_{\mathrm{S}}$ with a resulting observable ratio of $\left(0.55: 0.17: 0.28\right)_{\oplus\!}$.
There are also possible scenarios involving muon damping, where the highly magnetized source environment leads to a strong cooling of the comparatively long-lived muon through synchrotron emission before it decays~\cite{Kashti:2005qa,PhysRevD.75.123005}. As a result, only one high-energy $\nu_{\mu}$ is produced overall from the charged pion decay, $\left(0:1:0\right)_{\mathrm{S}}$,
which after oscillations would be observed as $\left(0.17:0.47:0.36 \right)_{\oplus\!}$.

High energy neutrinos of astrophysical origin have been measured with the IceCube Neutrino Observatory since 2013~\citep{IceCube:2013_HE,PRL_PeV,IceCube:2014_3HE, HESE7.5, IceCube:2015qii, IceCube:2016umi, IceCube:2018pgc, NorthernTracks, SBUCascades, MESE_2yr, IceCube:2024fxo, IceCube:2020fpi}. IceCube detects neutrinos of all flavors using its 5160 digital optical modules (DOMs) mounted on strings embedded in the Antarctic ice~\cite{Aartsen:2016nxy}. 
IceCube studies cosmic neutrinos by detecting Cherenkov light from the charged products of neutrino interactions. 
Charged current (CC) deep inelastic scattering (DIS) from $\nu_e$, $\nu_{\mu}$, and $\nu_{\tau}$, where the neutrinos scatter off the nucleons in ice producing the corresponding charged lepton and hadrons, result in characteristic morphologies within the detector. CC interactions of $\nu_{e}$ result in showers of particles, known as `cascades'. Cascades are also produced when neutrinos undergo neutral current (NC) DIS, where only the outgoing hadrons deposit energy in the detector volume. 
`Tracks' are generated via $\nu_{\mu}$ CC DIS, where the outgoing muon travels for several kilometers at TeV energies and deposits energy along its path. 
Tracks are also formed in $\nu_{\tau}$ CC DIS, if the outgoing $\tau$ lepton further decays into a muon. This occurs with a branching fraction of $\sim\,17\%$, and is hard to distinguish from tracks generated by $\nu_{\mu}$ CC interactions. 
Tau leptons from $\nu_{\tau}$ CC interactions with $E_{\tau}\gg 1$\,PeV can also generate tracks inside the detector volume if they decay outside the detector. 
`Double cascades' are produced by $\nu_{\tau}$ CC DIS where the initial hadronic energy from the DIS is visible as a first cascade and the resulting $\tau$~lepton travels at least $\mathcal{O}(10\,\mathrm{m})$ and decays (into electrons $\sim\,17.8\%$ of the time, and hadrons $\sim\,64.8\%$).
The mean decay length of the $\tau$~lepton scales with its energy, and can be approximated as $\langle L \rangle\propto 50~\mathrm{m}\,(E_{\tau}/\mathrm{PeV})$~\cite{IceCube:2015vkp}. Double cascades are hard to identify with IceCube since the spacing between the DOMs ($125$\,m horizontal spacing  and 17\,m vertical spacing~\cite{Aartsen:2016nxy}) is generally larger than the $\tau$ decay length. 
However, the timing information recorded by the DOMs can aid in resolving these events.
Initial evidence for $\nu_{\tau}$ was reported by the IceCube collaboration for two very energetic $\nu_{\tau}$ candidates that produced well-separated double cascades~\cite{IceCube:2020fpi}. Subsequently, a high-significance detection of $\nu_{\tau}$ was obtained with a lower energy threshold, leveraging subtle signatures produced by two cascades with lower separation distance~\cite{IceCube:2024nhk}.

IceCube has previously measured the flavor ratio of cosmic neutrinos with samples of cascades and tracks~\citep{IceCube:2015rro, icecube_collaboration_combined_2015, IceCube:2018pgc} and with a high energy sample of all three morphologies~\cite{IceCube:2020fpi}.
Here, we present results of the measurement of the astrophysical neutrino flavor ratio with 11.4 years of IceCube data, where events have energies ranging from 1\,TeV to more than 10\,PeV and are classified as cascades, tracks or double cascades.
The study is conducted using a sample of Medium Energy Starting Events (MESE), which utilizes events with vertices contained inside the detector volume ('starting events'). The MESE sample has recently provided evidence for a change in the spectral index of the diffuse astrophysical neutrino flux~\citep{BreakPRL,BreakPRD}. The study presented here identifies double-cascade events in the MESE sample, in addition to the existing cascade and track classification in~\cite{BreakPRL}. This facilitates the presented flavor-ratio measurement by helping to break the degeneracy between $\nu_e$ and $\nu_{\tau}$ events.

\textbf{Data sample:}
The major backgrounds for detecting astrophysical neutrinos are atmospheric muons and neutrinos that arise from cosmic-ray air showers. 
Background events from the Northern Hemisphere consist of atmospheric neutrinos, while in the southern sky, atmospheric muons are able to penetrate the ice and trigger the detector at a rate of $\sim3$\,kHz. 
The MESE dataset is built upon the concept of using veto regions to reject background muons entering the detector and in turn retain starting events.
Selecting starting events at lower $E_\nu$ threshold compared to~\cite{IceCube:2013_HE} allows the dataset to have enhanced sensitivity to neutrinos of all flavors, from the entire sky.
An initial version of the MESE sample was previously used to evaluate the astrophysical flux using 2 years of IceCube data~\cite{MESE_2yr}. The updated MESE sample makes use of improved selection methods, event reconstructions, and treatment of systematic uncertainties~\cite{BreakPRD}.
All simulations used in this analysis are based on the in-ice light propagation model described in~\citep{2013NIMPA.711...73A, Chirkin:2013lpu}. Observed event properties are also reconstructed based on this ice model. Further details regarding the MESE selection procedure can be found in~\cite{BreakPRD}. Compared to~\cite{BreakPRL} 
31 additional events are included here, since a further cut requiring 5035 active DOMs is applied in~\cite{BreakPRL}.

\textbf{Event classification:}
The events within the data sample are classified into tracks and cascades during the selection procedure described in~\citep{BreakPRL,BreakPRD}, and reconstruction algorithms are applied according to their classification. 
An additional classification of double cascades is performed for the flavor analysis reported here. 
The cascades/tracks classification is performed with a Deep Neural Network (DNN) which was trained to distinguish five event classes~\citep{DNN_TheoGlauch,dissertationTheo}. 
The DNN identifies features in the event morphology based on timing and charge information from the recorded signal of the event. It assigns a score to each event type, based on which we classify events as starting cascades or starting tracks in the sample. 
The classification power of the DNN increases with energy, and it is estimated from simulation to correctly identify $\sim\,88\%$ of starting cascades 
and $\sim\,97\%$ of starting tracks  
above 1 TeV.

The DNN classifies $\sim$ 82\% of true double-cascade events as cascades, since it is not explicitly trained on this event class. Previous searches with IceCube's high-energy data sample utilized a likelihood-based classification of double-cascade events~\cite{IceCube:2020fpi}. 
A convolutional neural network was used for recent observations of $\tau$ neutrinos leaving nuanced signatures in the DOMs~\cite{IceCube:2024nhk}. Here, we use the strategy from~\cite{IceCube:2020fpi} to select double cascade events, as it was already established during this analysis's development. All events that pass the final level MESE selection, already classified as cascades or tracks, undergo a maximum-likelihood based reconstruction under the double-cascade hypothesis. 
The fit utilizes the timing and spatial information of the event to reconstruct the following parameters: the energies of the two cascades ($E_1,\,E_2$), the spatial separation between them ($L$, directly associated to the $\tau$ decay length), and the direction and vertex of the event. 
The reconstructed total deposited energy of the event ($E_{\mathrm{tot}}$) is obtained from an algorithm that determines an event's unfolded energy along several segments in its path~\cite{IceCube:2013dkx}.
Events are classified as double cascades based on reconstructed quantities obtained from this fit. We define cuts based on the energy asymmetry $A_E=(E_1-E_2)/(E_1+E_2)$
and the energy confinement $E_C=(E_{C,1} + E_{C,2})/E_{\mathrm{tot}}$, where $E_{C,i}$ represents the sum total of the energy deposited within 40~m from the respective vertex for each cascade $i=1,\,2$, of the event.\footnote[1]{For $E_C$, depositions within 40~m of either cascade are summed only once, avoiding double counting in overlapping regions. For true double cascades $E_{\mathrm{tot}} = E_{C,1} + E_{C,2} = E_1+E_2$ holds.}

Events with $E_C\ge 0.99$, $-0.98\le A_E \le 0.3$, $L\ge 10$~m and $E_{\mathrm{tot}} \ge 30$~TeV are selected as double cascades, as these cuts allow a rejection of $\sim99\%$ of true cascades and true tracks as determined from MC distributions. Events that do not pass these cuts are retained as cascades or tracks, depending on their original classification from the DNN. The selected double cascade events are expected to have a purity of $\sim70\%$, based on simulations.
The simulated event expectation is shown in Tab.~\ref{tab:event_numbers}.

\textbf{Method:}
We use a forward folding binned likelihood method for conducting the flavor measurement using the \textsc{NNMFit} framework~\cite{BreakPRD}.
The analysis structure, the methods, and the parameters included in the fit reported here remain consistent with the MESE analysis described in~\citep{BreakPRL,BreakPRD}.
We split the data events according to the three morphologies: cascades, tracks, and double cascades, each binned in their respective observable space ($E_{\mathrm{reco}}$, cos($\theta_{\mathrm{reco}}$)), ($E_{\mathrm{reco}}$, cos($\theta_{\mathrm{reco}}$)) and ($E_{\mathrm{reco}}$, cos($\theta_{\mathrm{reco}}$), $L_{\mathrm{reco}}$) where $\theta_{\mathrm{reco}}$ is the reconstructed zenith angle.
The reconstructed energy~($E_{\mathrm{reco}}$) binning used for each morphology is shown in Tab.~\ref{tab:Bins2}. We use 10 bins for the full range of $\cos({\theta_\mathrm{reco}})$ for all morphologies.
The length observable ($L_{\mathrm{reco}}$) between the two cascades of a $\nu_{\tau}$ event is divided into 10 bins of $\log(L_{\mathrm{reco}})$ between 10-1000\,m.

\begin{table}[h]
\setlength{\tabcolsep}{4pt}
\renewcommand{\arraystretch}{1.1}
\caption{Binning used for the reconstructed energy of each type of morphology. The bins are logarithmic.}
\begin{tabular}{c|c|c}
\textbf{Morphology} & \textbf{Energy range (GeV)} & \textbf{No.\,of bins} \\ \hline
Cascades            & $10^3-10^7$                 & 22                   \\ \hline
Tracks              & $10^3-10^7$                 & 13                   \\ \hline
Double cascades     & $3\times10^4-10^7$          & 13                  
\end{tabular}
\label{tab:Bins2}
\end{table}
We perform a maximum-likelihood ($\mathcal{L}_{\mathrm{max}}$) fit of the independent event classes simultaneously (in 2D for cascades and tracks, and 3D for double cascades), where they have shared parameters.
The fit parameters include the astrophysical neutrino flux and flavor ratio, the conventional atmospheric neutrino flux from decay of pions and kaons in cosmic-ray air showers, the prompt atmospheric neutrino flux from decay of charmed hadrons, and the atmospheric muon background. We also include systematic uncertainties as nuisance parameters in the fit, which affects the total predicted number of events. See~\cite{BreakPRD} for a comprehensive description of the nuisance parameters.

Neutrinos of all flavors are simulated, with NuGen~\cite{NuGen}, followed by the detector response, with a nominal flavor ratio of 1:\,1:\,1. Their respective fraction can be modified by reweighting the simulations during the $\mathcal{L}_{\mathrm{max}}$ fit. Details of the simulations are discussed in~\cite{BreakPRD}.
The atmospheric flux components used for the forward-folding fit are derived from model predictions. We assume Gaisser H4a as the primary CR composition model~\cite{gaisser_spectrum_2012} and Sibyll 2.3c as the hadronic interaction model~\cite{riehn_hadronic_2018}. 
The corresponding atmospheric neutrino fluxes, conventional flux and prompt flux, are derived using the package \textsc{MCEq}~\cite{fedynitch_calculation_2015,noauthor_mceq-projectmceq_2023}. These flux predictions are applied as weights to simulated events.
Modifications to the energy distribution of tau leptons generated by CC DIS due to their polarization~\cite{Garg:2022ugd} are implemented as corrections to the weights of the baseline simulations.
The atmospheric muon background is modeled based on a template from simulations of single muons described in~\cite{MuonGun} that has been smoothed using a kernel density estimator.
Variations to the nominal predictions of these atmospheric flux contributions and their uncertainties  are handled via nuisance parameters included in the $\mathcal{L}_{\mathrm{max}}$ fit as discussed in~\cite{BreakPRL}.
Modifications to the atmospheric neutrino flux caused by the self-veto effect, by which an atmospheric neutrino is removed due to an accompanying muon from the same air shower~\cite{PhysRevD.90.023009},
is included as a nuisance parameter in the fit as well. 
A global parameter further allows for changes of the inelasticity distribution of CC and NC neutrino interactions (derived from CSMS~\cite{Cooper-Sarkar:2011jtt}), across the entire energy range. 
These parameterizations are the same as those used in~\citep{BreakPRL,BreakPRD}.

Detector-related systematics mainly pertain to the uncertainties in our knowledge of the ice and the optical efficiency of the DOMs. Parameters that account for the absorption and scattering of light in the bulk ice, and the anisotropy of the ice which causes asymmetric propagation of light~\cite{Chirkin:2013lpu} are included in the fit. Global “hole-ice” parameters, which modify the angular photon acceptance of the DOMs to
account for the different scattering of photons along the refrozen column of ice surrounding the DOMs, are also incorporated. We include a global parameter that accounts for variations in the mean photon detection efficiency of the DOMs.
These detector systematics are included via the method described in~\cite{aartsen_efficient_2019,ganster_combined_2022}, where perturbations to the nominal values of the above-mentioned parameters are generated per simulated event. 
We perform the flavor measurement assuming the astrophysical flux model to be a broken power law (BPL)\footnote[2]{The broken power law flux is modeled as $\Phi_{\nu+\bar{\nu}} = \phi_0 \left( \frac{E_{\nu}}{E_{\mathrm{break}}} \right)^{-\gamma} \left( \frac{E_{\mathrm{break}}}{100\,\mathrm{TeV}} \right)^{-\gamma'}$, with $\gamma=\gamma_1$ ($E_{\nu}<E_{\mathrm{break}}$), $\gamma=\gamma_2$ ($E_{\nu}>E_{\mathrm{break}}$); and $\gamma'=\gamma_1$ if $E_{\mathrm{break}}>100\,\mathrm{TeV}$, else $\gamma_2$.}, 
remaining consistent with the best-fit model from~\cite{BreakPRL}, which uses the MESE sample.
An accurate modeling of the spectral shape is essential for the flavor measurement to avoid bias in interpretation, as shown in App.~\ref{app:SPL} where we additionally perform a cross-check assuming a single power law (SPL) as the astrophysical flux model.
Two parameters that account for the fractional contribution of $\nu_e$ and $\nu_{\tau}$ to the overall flux normalization (with $f_e+\,f_{\mu}+\,f_{\tau}\,=\,1$) are the main physics parameters we measure here. We assume that the same flavor ratio holds across all measured energies and leverage the identifiable high energy double-cascade events within the sample to constrain the $\nu_{\tau}$ fraction.

\textbf{Results:}
\label{sec:results}
We report the best fit for the flavor composition of the astrophysical neutrino flux, when performing the $\mathcal{L}_{\mathrm{max}}$ fit under the assumption of the BPL model.
Tab.~\ref{tab:event_numbers} compares the observed data to model predictions (see App.~\ref{app:data_mc_comparison} for more details). 
\begin{table}[h!]
\renewcommand{\arraystretch}{1.1}
\caption {Event counts in data compared to MC predictions of the best-fit for BPL and SPL models. The predictions and observed counts are compatible within expectations Poisson from fluctuations.}
\begin{tabular}{c|cl|cl|c}
&
\multicolumn{2}{c|}{\begin{tabular}[c]{@{}c@{}}Cascades\\ ($E$ \textgreater \,1~TeV)\end{tabular}} &
\multicolumn{2}{c|}{\begin{tabular}[c]{@{}c@{}}Tracks\\ ($E$ \textgreater \,1~TeV)\end{tabular}} &
  \begin{tabular}[c]{@{}c@{}}Double Cascades\\ ($E$ \textgreater \,30~TeV)\end{tabular} \\ \hline
Data &  \multicolumn{2}{c|}{4960} &  \multicolumn{2}{c|}{4919} & 9 \\ \hline
BPL &  \multicolumn{2}{c|}{$4953.6\pm 154.6$} & \multicolumn{2}{c|}{$4876.2\pm 136.1$} & $7.0\pm 0.9$ \\ \hline
SPL & \multicolumn{2}{c|}{$4999.2\pm 160.4$} & \multicolumn{2}{c|}{$4825.4\pm 141.7$} & $9.1\pm 1.0$
\end{tabular}
\label{tab:event_numbers}
\vspace{-6mm}
\end{table}
\begin{figure}[h!]
\setlength{\abovecaptionskip}{-3pt}
\setlength{\belowcaptionskip}{-3pt}
\includegraphics[width=0.97\linewidth]{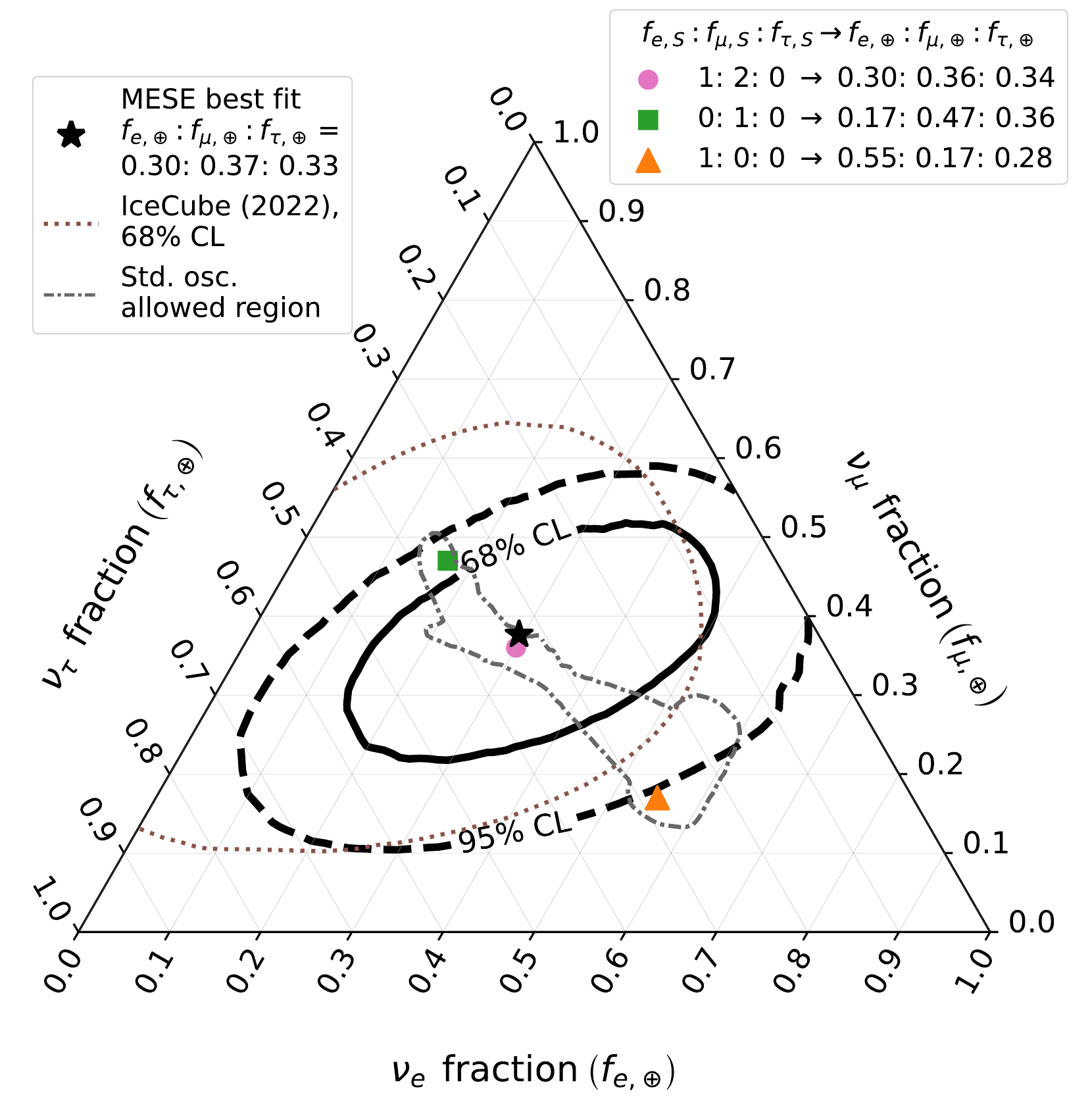}
\caption{\textbf{Ternary diagram of the results of the flavor-composition fit:} The axes show the fraction of $\nu_e$, $\nu_{\mu}$ and $\nu_{\tau}$ at Earth. 68\% and 95\% CL contours assuming the test statistic follows Wilks' theorem~\cite{Wilks} are shown as solid and dashed lines, respectively. Expected flavor composition at Earth, after standard oscillations, for benchmark production mechanisms (pion decay: circle, muon damping: square, neutron decay: triangle) and all possible flavor compositions after propagation (dash-dot line, from~\cite{Song:2020nfh}) are shown. The dotted line shows the 68\% CL contour from IceCube's last measurement~\cite{IceCube:2020fpi}.}
\label{fig:flav_bpl}
\end{figure} 
We classify 9 events as double cascades out of a total of 9888 events, consistent with expectations from MC. 

As shown in Fig.~\ref{fig:flav_bpl}, the 68\% confidence level (CL) contour closes for this fit, which is achieved for the first time with TeV--PeV astrophysical neutrinos.
The best fit ($\left(0.30 : 0.37 : 0.33\right)_{\oplus\!}$) is consistent with expectations of flavor ratios at Earth after the neutrinos undergo standard oscillations.
A dominant neutron-decay production mechanism is rejected with 95.3\% CL and
the fractions $f_{\tau,\oplus\!}$ = 0 and $f_{e,\oplus\!}$ = 0 are rejected with 91.9\% CL and 98.7\% CL, from Wilks' theorem–derived confidence regions of the $\mathcal{L}_{\mathrm{max}}$ fit. The validity of Wilk's theorem was tested using MC pseudoexperiments, as in~\cite{IceCube:2020fpi}.
The 95\% CL contour does not close along the $\nu_{\tau}$ axis. This can be attributed to the steep high-energy spectral index of the BPL model, for which the best-fit spectral parameters are $\phi_0\,=\,2.72^{+0.95}_{-0.92}\times\rm{10^{-18}/GeV/cm^2/s/sr}$, $\gamma_1\,=\,1.76^{+0.36}_{-0.26}$, $\gamma_2\,=\,2.81^{+0.076}_{-0.12}$ and $\log_{10}(E_{\mathrm{break}}/\mathrm{GeV})\,=\,4.5^{+0.12}_{-0.09}$.
These fit results are consistent with those reported in~\cite{BreakPRL}. 
A steep $\gamma_2$ results in a low number of $\nu_{\tau}$ events with identifiable double cascade morphologies (which is mostly comprised of events of the highest energies) as shown in Tab.~\ref{tab:event_numbers}.

\begin{figure}[h!]
\setlength{\abovecaptionskip}{-8pt}
\setlength{\belowcaptionskip}{-18pt}
\includegraphics[width=0.99\linewidth]{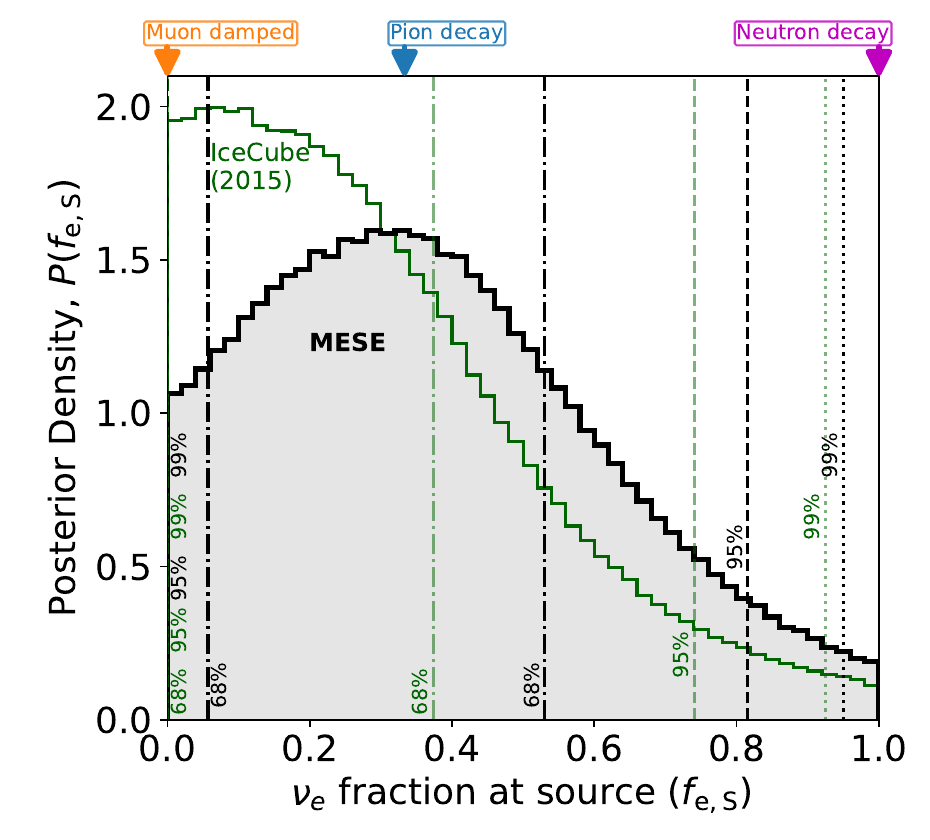}
\caption{\textbf{Posterior distribution of flavor composition at source:} Assuming $f_{\tau,\mathrm{S}}\,=0\,$ and global best-fit mixing parameters (normal ordering) with their $\chi^2$ distributions from~\cite{Esteban:2024eli}, and the likelihood profile obtained with the $\mathcal{L}_{\mathrm{max}}$ fit to the MESE dataset. The same with the likelihood profile from the IceCube (2015) analysis in~\cite{icecube_collaboration_combined_2015}, which reported more constraining rejection of neutron decay with the $\mathcal{L}_{\mathrm{max}}$ fit.}
\label{fig:source_post}
\end{figure}

This measurement is further used to constrain the flavor composition at source. Given the global measurements of neutrino-oscillation parameters and the flavor composition at Earth reported here, we derive a posterior distribution of the composition at source as in~\cite{Bustamante:2019sdb}.
We assess the probability distribution of $f_{\mathrm{e,S}}$ as shown in Fig.~\ref{fig:source_post}, with the assumption that no $\nu_{\tau}$ are produced at the sources, as expected from the benchmark scenarios, while ensuring that the flavor ratios add up to 1. The oscillation parameters are obtained from NuFit 6.0~\citep{Esteban:2024eli, Nufit6.0}. 
We observe a distribution that peaks towards the pion decay scenario and rejects the neutron decay scenario, which lies outside the 99\% confidence interval (CI). 
Although the 68\% CI favors pion decay, the muon-damped case is well within the 99\% CI, making it hard to exclude this case. However, in both cases, neutrinos will be produced by pion beams arising from $pp$ or $p\gamma$ interactions in the source.
Fig.~\ref{fig:source_post} also shows the posterior distribution obtained with a previous IceCube measurement~\cite{icecube_collaboration_combined_2015} using tracks and cascades, where the confidence region from the $\mathcal{L}_{\mathrm{max}}$ fit rejected neutron decay at sources with 99\% CL. The corresponding 99\% CI is marginally more constraining on the neutron decay scenario, and the distribution is skewed more towards the muon-damped case. However, this measurement had no distinction between double cascades and other morphologies, and was unable to break the $\nu_e-\nu_\tau$ degeneracy. 
The CI's with the MESE dataset is shown in Tab.~\ref{tab:confidence_intervals}.

\begin{table}[h!]
\renewcommand{\arraystretch}{1.1}
\caption {Bayesian CIs for $f_{e,\mathrm{S}}$ with normal and inverted ordering (without Super Kamiokande~\citep{Nufit6.0}), constructed as highest probability density intervals of the posterior distributions.}
\begin{tabular}{c|cl|cl|c}
&
\multicolumn{2}{c|}{68\% CI} &
\multicolumn{2}{c|}{{95\% CI}} &
  {99\% CI} \\ \hline
Normal Ordering &  \multicolumn{2}{c|}{(0.06, 0.53)} &  \multicolumn{2}{c|}{(0, 0.82)} & (0, 0.95) \\ \hline
Inverted Ordering &  \multicolumn{2}{c|}{(0.05, 0.52)} & \multicolumn{2}{c|}{(0, 0.82)} & (0, 0.95) \\ 
\end{tabular}
\label{tab:confidence_intervals}
\end{table}

\textbf{Discussion and Conclusion:} 
We report a measurement of the flavor ratio of cosmic neutrinos with the IceCube detector.
The best fit of the flavor ratio at Earth lies within the dash-dotted curve in Fig.~\ref{fig:flav_bpl}. Any outcome of the measurement which is not positioned within this region is inconsistent with the standard theory of oscillations, since all possibilities for the  initial flavor compositions at astrophysical sources end up here. The rest of the diagram cannot be reached within the standard 3-flavor scenario~\cite{Song:2020nfh}.
The improvements in the flavor measurement reported here, compared to previous measurements 
originate from the inclusion of TeV-scale cascades and tracks with the MESE sample, which form a larger fraction of the data than PeV events,  in combination with the additional identification of double cascade events. Our previous analysis that included all three morphologies~\cite{IceCube:2020fpi} utilized only the highest energy events, and therefore probed a different scale of propagation length per unit energy on average. The much larger event number from including lower-energy events, along with the improved identification using the DNN, enhances the sensitivity of the present measurement. 
Using cascade/track classification alone leaves the $f_e$–$f_\tau$ degeneracy unresolved, as in~\cite{icecube_collaboration_combined_2015}.
Including the double-cascade selection, despite the low statistics, yields superior constraints along the $f_e - f_{\tau}$ axis.
The results presented here provide constraints crucial for identifying the neutrino production mechanism at astrophysical sources. The Bayesian posterior analysis shown in Fig.~\ref{fig:source_post} provides a strong rejection of the neutron decay scenario, placing it outside the 99\%~CI.
It is also possible that the astrophysical flavor ratio varies with energy, owing to different neutrino production mechanisms at different energies. A test of this would require an energy-dependent flavor measurement, which is currently challenging because of the low statistics of double-cascade events. Future measurements that include enhanced identification of $\nu_{\tau}$ events as in~\cite{IceCube:2024nhk} or identification of $\nu_{\tau}$ tracks based on their energy deposition, which differs from muon tracks~\cite{IceCube:2018pgc} can potentially improve the statistics of $\nu_{\tau}$ events. 
The combined sample of exclusive $\tau$ neutrino candidates from this and other recent IceCube analyses~\citep{IceCube:2020fpi,IceCube:2024nhk} is the world's largest. Only one candidate is common between these analyses~\citep{IceCube:2020fpi,IceCube:2024nhk}. Given the low overall selection efficiencies for $\nu_\tau$, the two strategies are expected to yield largely disjoint event samples.

We assume equal production of neutrinos and antineutrinos in the measurement reported here. Currently, IceCube is not able to explicitly differentiate between $\nu$ and $\bar{\nu}$ signatures. This separation can be done via the identification of Glashow resonance events, which is a phenomenon that occurs only for $\bar{\nu}_e$ at the highest energies~\cite{IceCube:2021rpz}, thereby limiting the available statistics. Additionally, ongoing studies aim to discriminate $\nu$ and $\bar{\nu}$ in starting tracks by utilizing their differing inelasticity distributions, driven by their distinct valence-quark couplings~\cite{Skrzypek:2025tmg}.
Including separate treatments for $\nu$ and $\bar{\nu}$ with sufficient statistics can help constrain the flavor ratio further.
Future measurements with IceCube-Gen2~\cite{IceCube-Gen2:2020qha} will benefit from its larger collection volume enabling high-precision probes of the neutrino flavor ratio~\cite{Gen2:TDR}, including that of cosmic neutrinos at ultra-high energies.
\begin{acknowledgements}
The IceCube collaboration acknowledges the significant contributions to this manuscript from Aswathi Balagopal V. and Vedant Basu. We thank Mauricio Bustamante for valuable discussions.
The authors gratefully acknowledge the support from the following agencies and institutions:
USA {\textendash} U.S. National Science Foundation-Office of Polar Programs,
U.S. National Science Foundation-Physics Division,
U.S. National Science Foundation-EPSCoR,
U.S. National Science Foundation-Office of Advanced Cyberinfrastructure,
Wisconsin Alumni Research Foundation,
Center for High Throughput Computing (CHTC) at the University of Wisconsin{\textendash}Madison,
Open Science Grid (OSG),
Partnership to Advance Throughput Computing (PATh),
Advanced Cyberinfrastructure Coordination Ecosystem: Services {\&} Support (ACCESS),
Frontera and Ranch computing project at the Texas Advanced Computing Center,
U.S. Department of Energy-National Energy Research Scientific Computing Center,
Particle astrophysics research computing center at the University of Maryland,
Institute for Cyber-Enabled Research at Michigan State University,
Astroparticle physics computational facility at Marquette University,
NVIDIA Corporation,
and Google Cloud Platform;
Belgium {\textendash} Funds for Scientific Research (FRS-FNRS and FWO),
FWO Odysseus and Big Science programmes,
and Belgian Federal Science Policy Office (Belspo);
Germany {\textendash} Bundesministerium f{\"u}r Bildung und Forschung (BMBF),
Deutsche Forschungsgemeinschaft (DFG),
Helmholtz Alliance for Astroparticle Physics (HAP),
Initiative and Networking Fund of the Helmholtz Association,
Deutsches Elektronen Synchrotron (DESY),
and High Performance Computing cluster of the RWTH Aachen;
Sweden {\textendash} Swedish Research Council,
Swedish Polar Research Secretariat,
Swedish National Infrastructure for Computing (SNIC),
and Knut and Alice Wallenberg Foundation;
European Union {\textendash} EGI Advanced Computing for research;
Australia {\textendash} Australian Research Council;
Canada {\textendash} Natural Sciences and Engineering Research Council of Canada,
Calcul Qu{\'e}bec, Compute Ontario, Canada Foundation for Innovation, WestGrid, and Digital Research Alliance of Canada;
Denmark {\textendash} Villum Fonden, Carlsberg Foundation, and European Commission;
New Zealand {\textendash} Marsden Fund;
Japan {\textendash} Japan Society for Promotion of Science (JSPS)
and Institute for Global Prominent Research (IGPR) of Chiba University;
Korea {\textendash} National Research Foundation of Korea (NRF);
Switzerland {\textendash} Swiss National Science Foundation (SNSF).
\end{acknowledgements}

%
%
\newpage

\appendix
\section{Data/MC comparison}
\label{app:data_mc_comparison}
\begin{figure*}[th!]
    \includegraphics[width=0.32\linewidth]{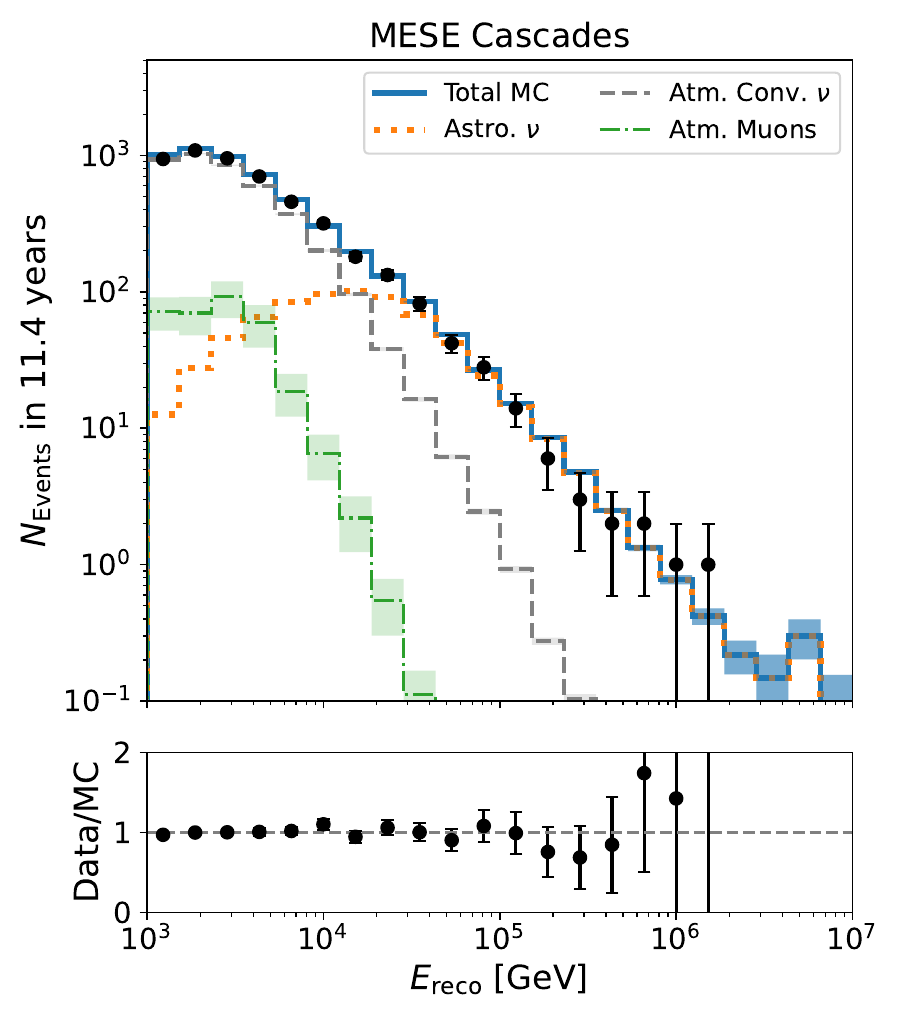}
    \includegraphics[width=0.32\linewidth]{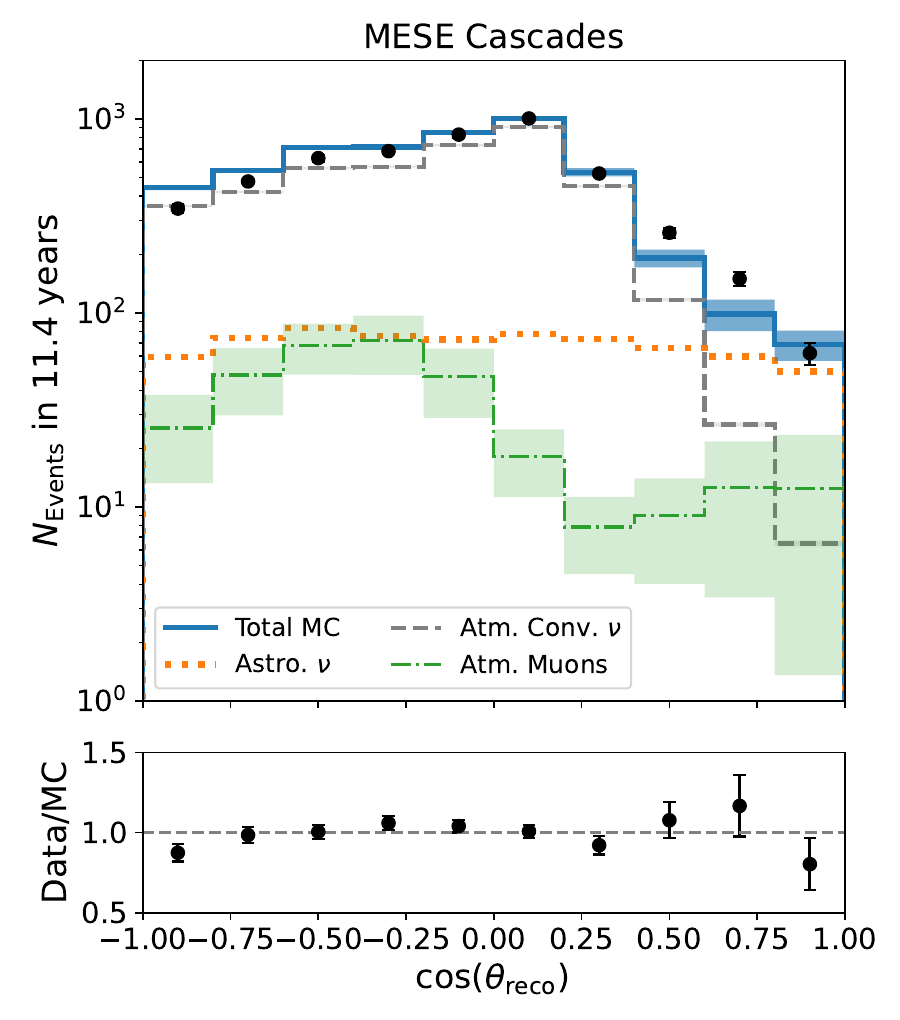}\\
    \includegraphics[width=0.32\linewidth]{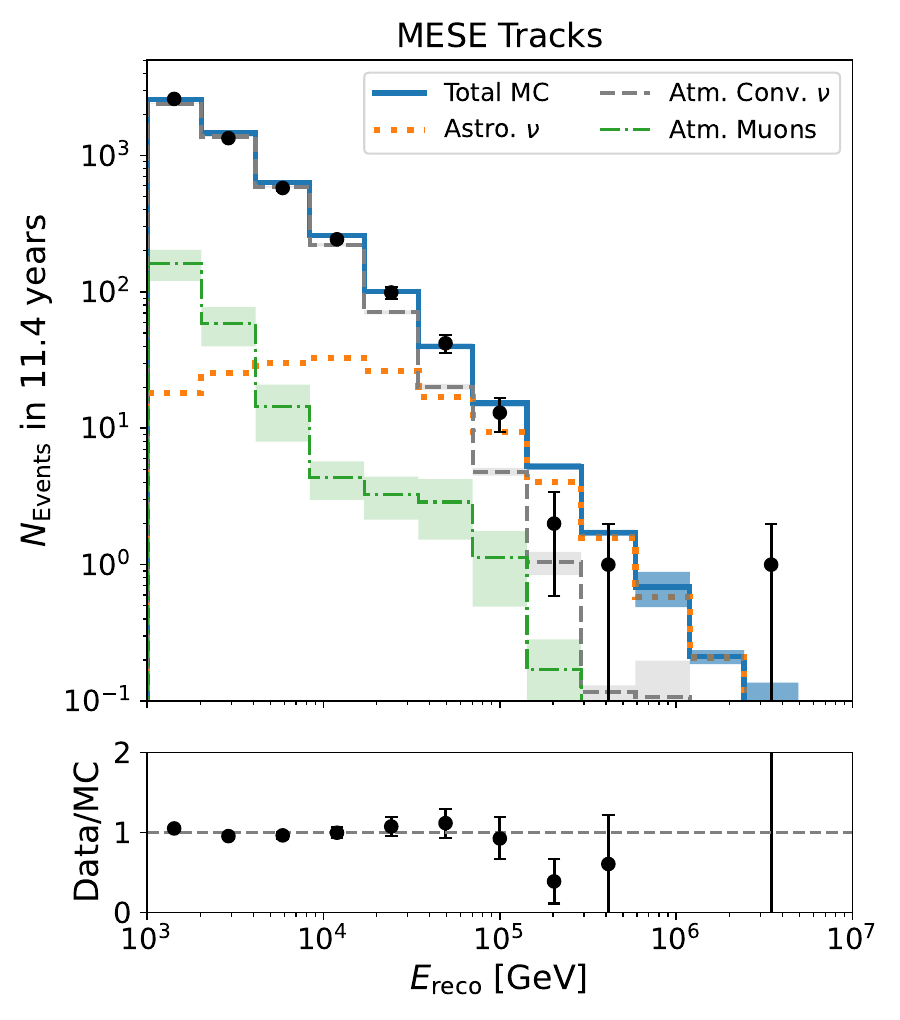}
    \includegraphics[width=0.32\linewidth]{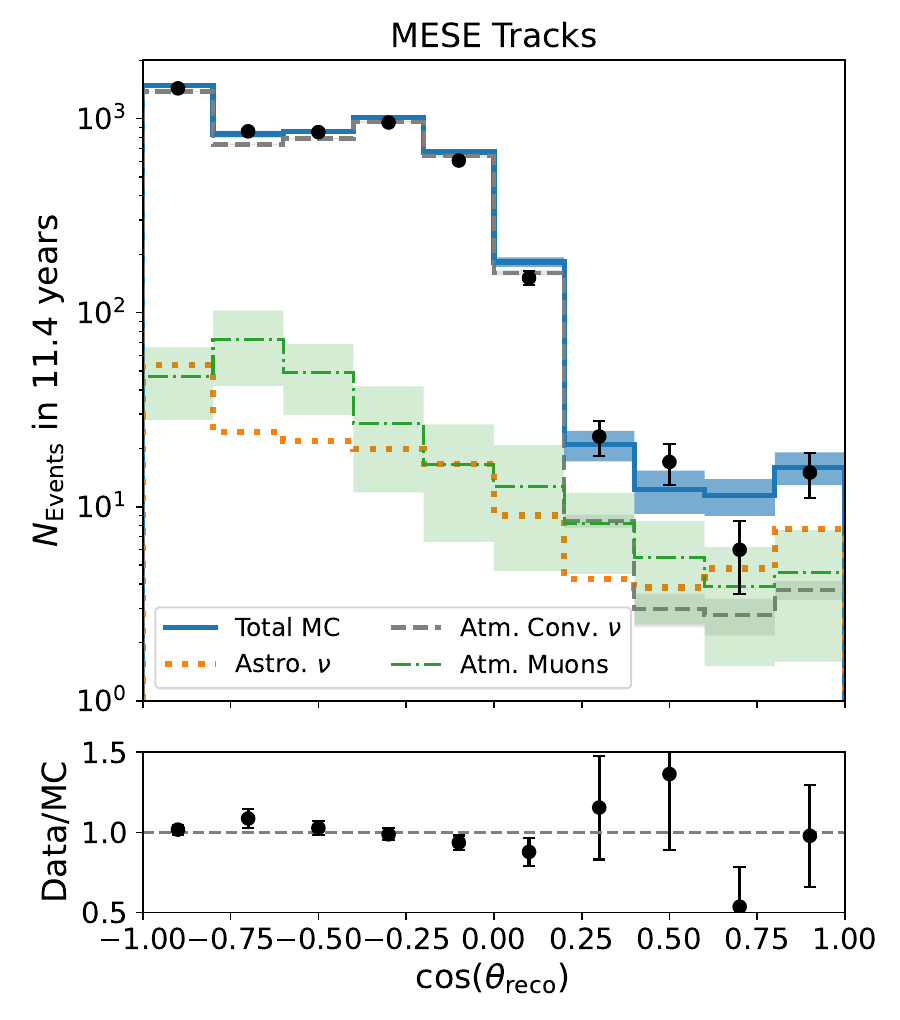}\\
    \includegraphics[width=0.32\linewidth]{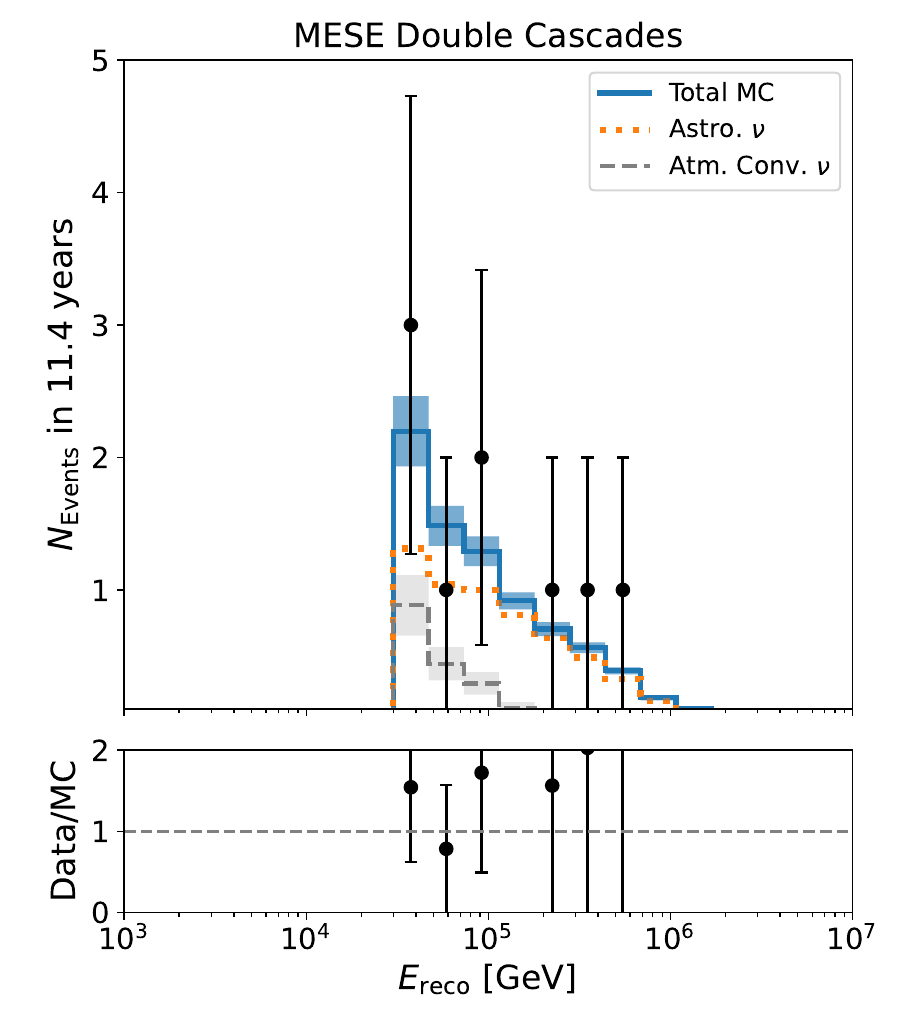}
    \includegraphics[width=0.32\linewidth]{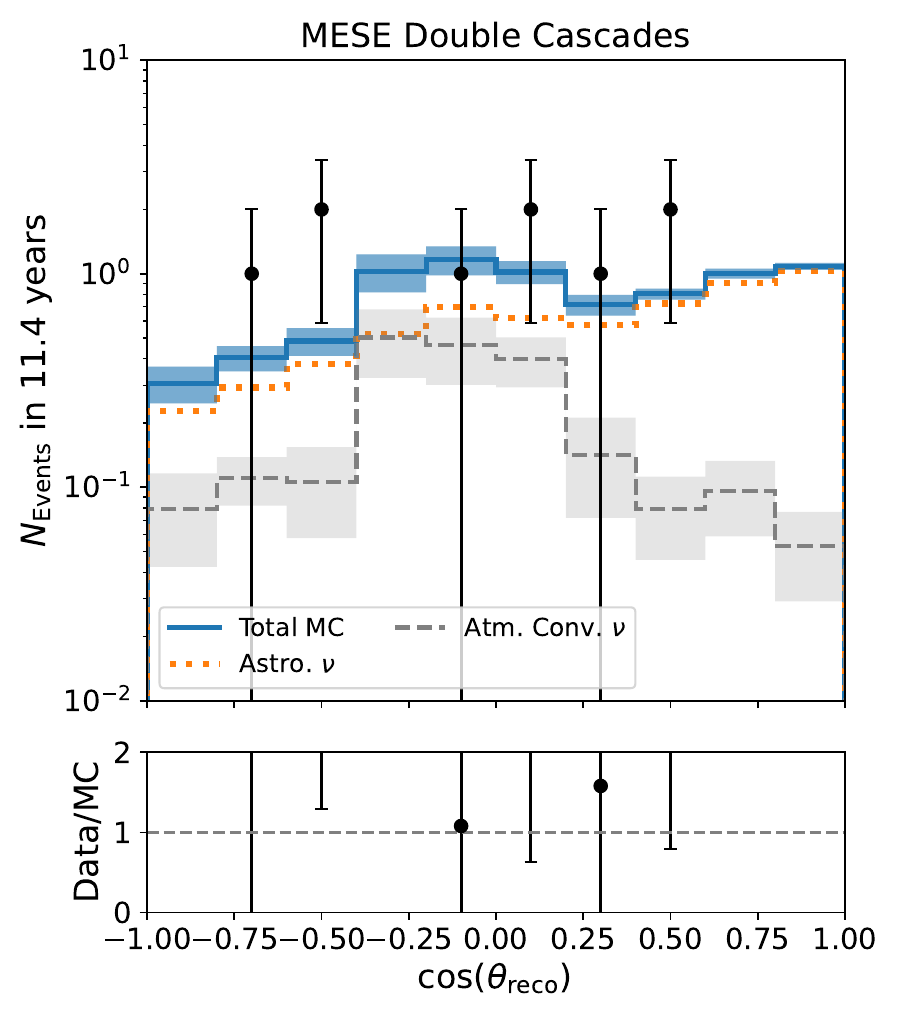}
    \includegraphics[width=0.32\linewidth]{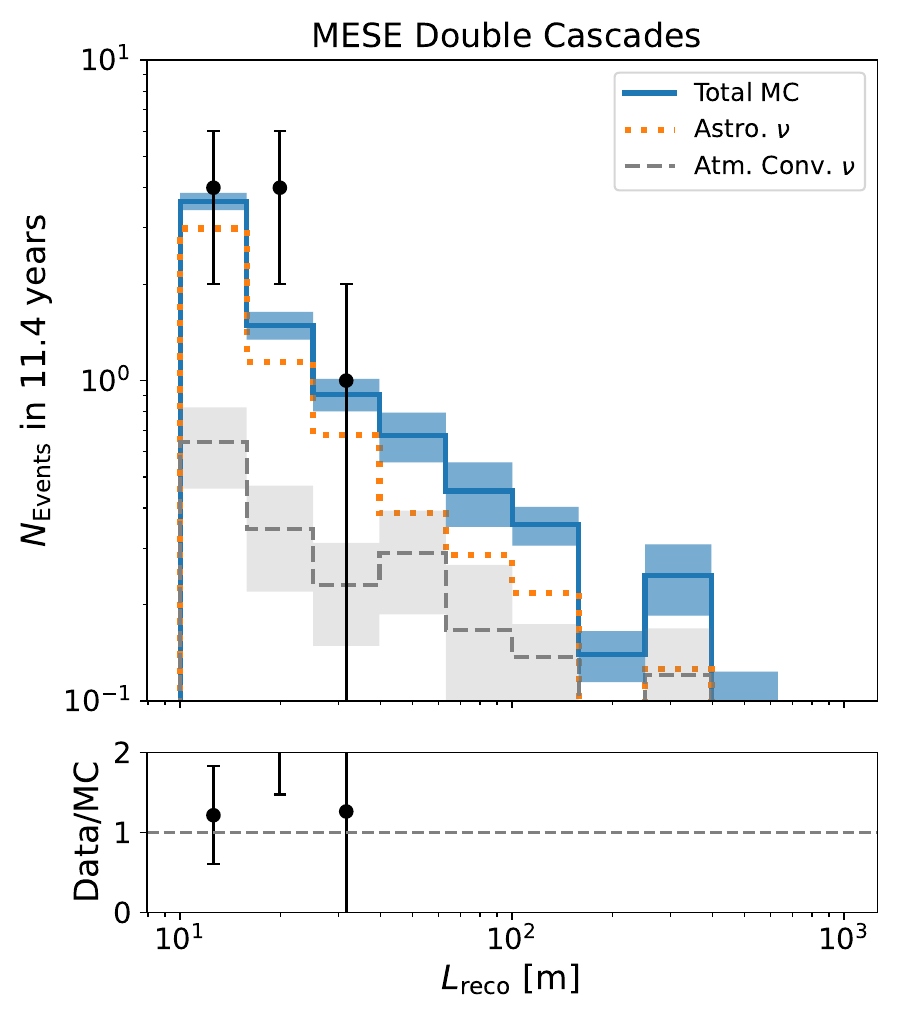}
    \caption{\textbf{Histograms of observables:} The set of 7 observables used for performing the flavor measurement, with 11.4 years of data. We assume a broken power law model for the astrophysical flux. We fit a prompt contribution of zero, and hence it is not shown in the figure. No muon component survives the double cascade selection procedure.}
    \label{fig:data-mc}
\end{figure*}

Figure~\ref{fig:data-mc} compares the observed data to MC predictions for the observable in each channel on which the fit is performed. The 1D projections are shown in the figure. The fit is performed in 2D for cascades and tracks and in 3D for double cascades. Data and the best-fit MC are compatible with each other within $2\,\sigma$ as shown by the ratio between them in the bottom panels of Fig.~\ref{fig:data-mc}. Table~\ref{tab:double_cascades} shows the reconstructed properties for the events classified as double cascades within the MESE sample.

\begin{table}[h]
\renewcommand{\arraystretch}{1.1}
\setlength{\abovecaptionskip}{2pt}
\setlength{\belowcaptionskip}{1pt}
\caption {Reconstructed properties of the events classified as double cascades.}
\begin{tabular}{c|c|c|c}
Event number & $E_1$, $E_2$ (TeV) & $E_{\mathrm{tot}}$ (TeV) & $L$ (m) \\ \hline\hline
\#1                   & 3.7, 236.2            & 239.9               & 14.3                \\ \hline
\#2                   & 6.9, 26.2             & 32.5                & 17.3                \\ \hline
\#3                   & 4.6, 325.8            & 330                 & 18.3                \\ \hline
\#4                   & 5.5, 77.2             & 82.3                & 21.4                \\ \hline
\#5                   & 20.6, 11.36           & 32                  & 12.2                \\ \hline
\#6                   & 32.5, 27.9            & 60.8                & 11.1                \\ \hline
\#7                   & 22.3, 28.9            & 45.7                & 39.8                \\ \hline
\#8                   & 47.4, 44.7            & 92                  & 16                  \\ \hline
\#9                   & 299.9, 191.5          & 491.3               & 10.6               
\end{tabular}
\label{tab:double_cascades}
\end{table}

\section{Comparison with an SPL fit}
A fit for the astrophysical flavor ratio where we assume an SPL model was performed as a cross-check, since previous IceCube measurements were consistent with an SPL~\cite{IceCube:2024fxo} unlike the latest measurement in~\cite{BreakPRL}. The best fit obtained with an SPL is $f_{e,\oplus\!}:f_{\mu,\oplus\!}:f_{\tau,\oplus\!}\,=\,0.28 : 0.36 : 0.36$, which is consistent with the best fit value of $f_{e,\oplus\!}:f_{\mu,\oplus\!}:f_{\tau,\oplus\!}\,=\,0.30 : 0.37 : 0.33$ obtained with the BPL model. The SPL fit allows the astrophysical flux normalization and the spectral index to be free parameters in the fit. The fit values at the best-fit point for these parameters are $\phi^{\nu+\bar{\nu}}\,=\,2.55$ and $\gamma\,=\,2.54^{+0.05}_{-0.04}$, compatible with the results reported in~\cite{BreakPRL} with the MESE sample, where a flavor ratio of 1:1:1 was assumed.

\begin{figure}[h]
\includegraphics[width=1\linewidth]{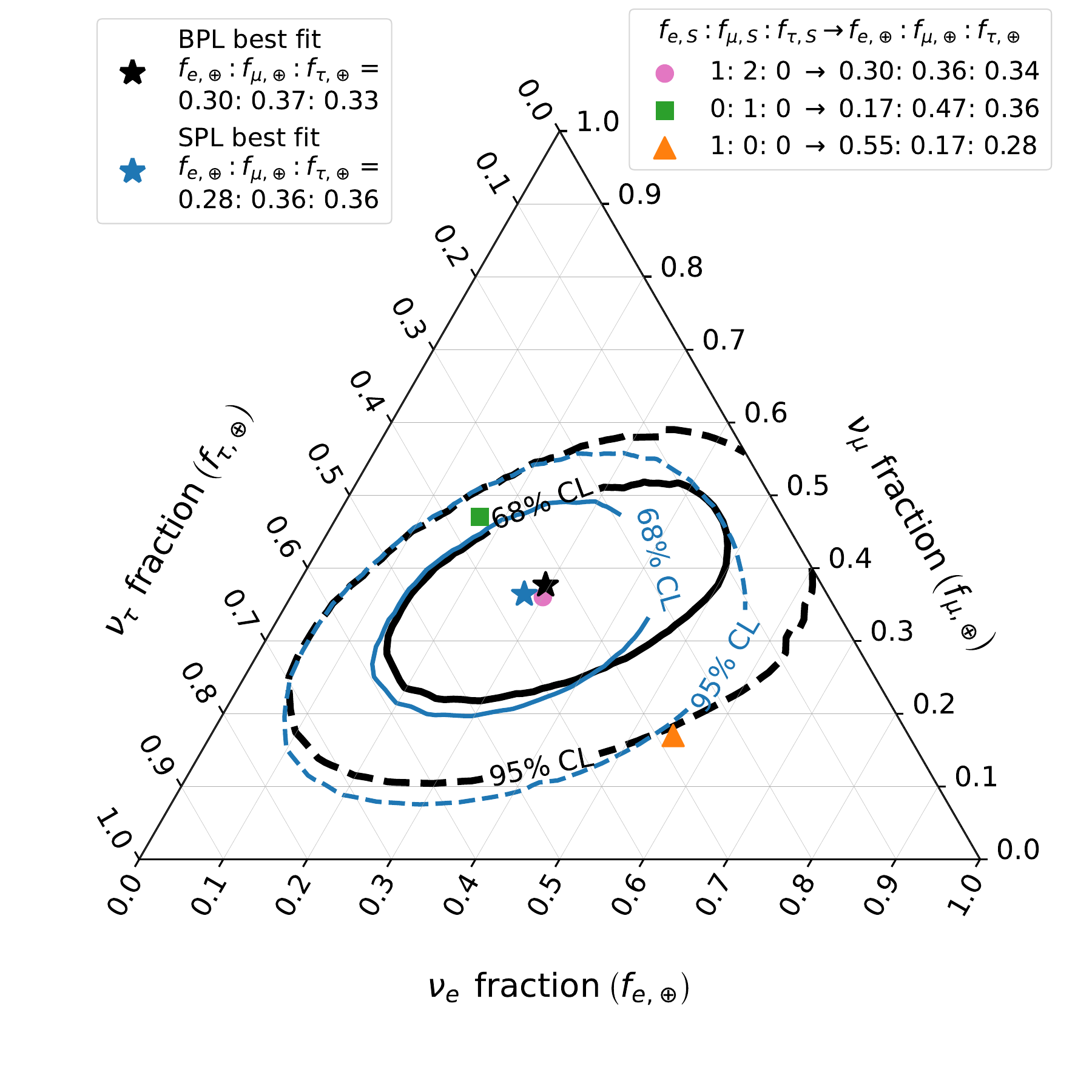}
\caption{\textbf{Ternary flavor composition for different spectral assumptions:} The baseline measurement, assuming a broken power law, is compared to a single power law assumption. The contour size shrinks along the $\nu_{\tau}$ axis for an SPL, since it predicts more tau neutrinos due to its harder spectrum at higher energies. The best fits remains close to each other for both models. Also $f_e$ and $f_{\mu}$ CLs remain similar.}
\label{fig:SPL_BPL}
\vspace{-7mm}
\end{figure}

Fig.~\ref{fig:SPL_BPL} compares the baseline fit with a BPL model with a fit where we assume the underlying model to be an SPL. The 68\% and 95\% CL contours for both models are shown. We see that these contours shrink along the $\nu_{\tau}$ axis for the SPL fit, and in particular, the 95\% contour is seen to close along $\nu_{\tau}\approx 0.1$. A stronger constraint on the $\nu_{\tau}$ fraction arises due to the harder spectral index for the SPL fit at higher energies, when compared to the soft index of $\sim$ 2.8 for the BPL fit. This results in more double cascade events being predicted for the SPL model than for the BPL model, as shown in Tab~\ref{tab:event_numbers}. The fit with the BPL model has a significant improvement in the likelihood ($-2\Delta\ln\mathcal{L} = 24$) when compared to the fit with the SPL model. This also demonstrates the importance of using the flux model that best describes the observed data to obtain accurate constraints on the astrophysical flavor ratio.
~\label{app:SPL}

\section{Flavor ratio at source}
~\label{sec:source_posterior}
\begin{figure}[h]
    \centering
    \includegraphics[width=1\linewidth]{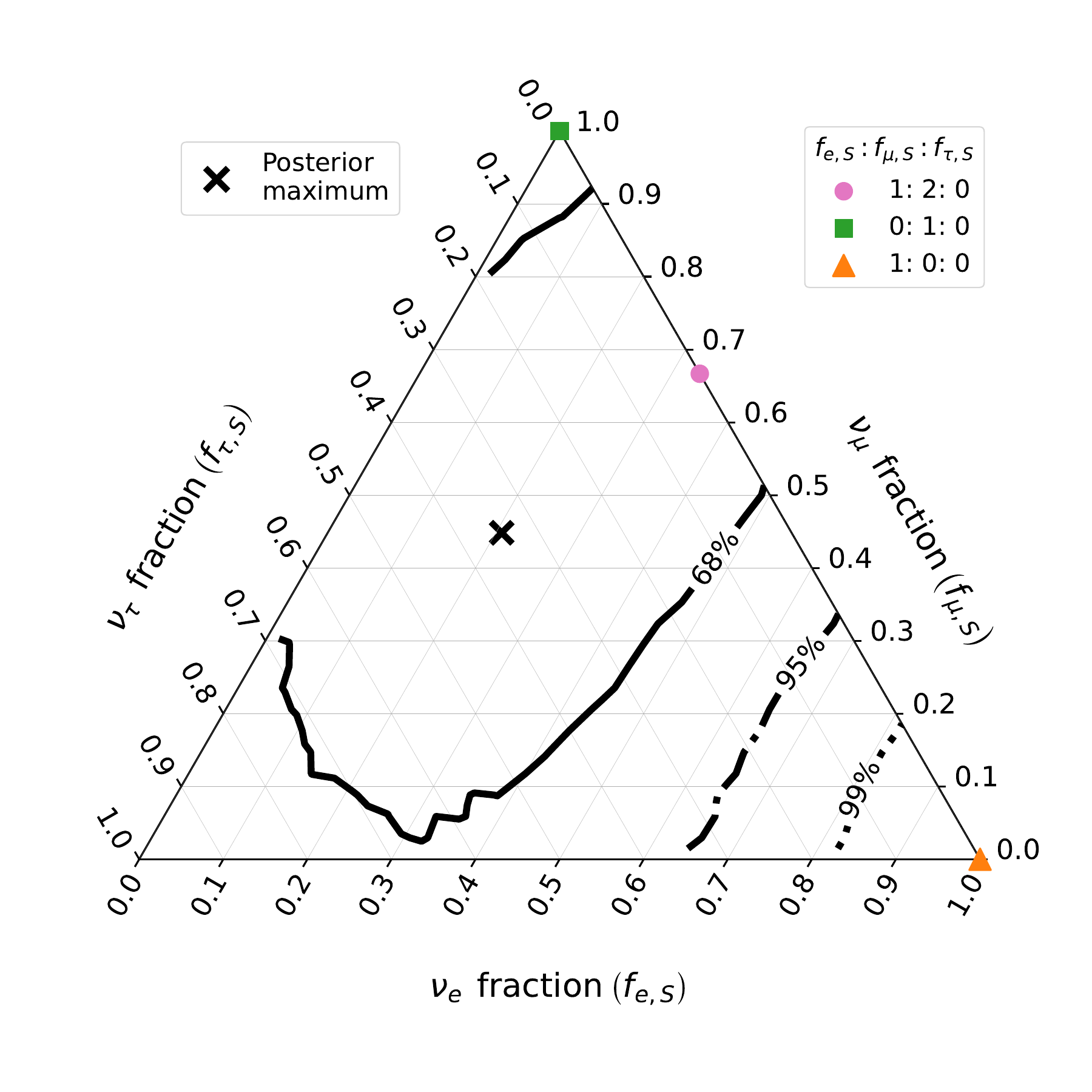}
    \caption{\textbf{Posterior at source, no prior on f$_{\tau,\,\rm{S}}$:} Here we do not assume the additional prior on the fraction of $\nu_{\tau}$ at source. The remaining assumptions are the same as that in Fig.~\ref{fig:source_post}. Contours describing the high probability density regions, constructed from the posterior distribution, are shown.}
    \label{fig:pos_no_tauprior}
\vspace{-7mm}
\end{figure}
An additional study is conducted on the Bayesian posterior at the source, where we make no assumptions on the composition at the source, and allow the fraction of tau neutrinos at the source to take any values between 0 and 1. The highest probability density contours derived from the posterior distribution are shown in Fig.~\ref{fig:pos_no_tauprior}. While the muon-damped scenario is outside the 68\% contour, it is within the 95\% contour. The neutron decay at source scenario is well rejected even with no prior on f$_{\tau,\,\rm{S}}$.
\section{Ice systematics}
The optical properties of the ice medium can impact the identification strength of double cascades. Effects like light scattering and absorption can distort the observed light from the two cascades and cause uncertainty in their measurement.  An anisotropy in the propagation of light in ice~\cite{2020JInst..15P6032A}, which results in a directional dependence of light propagation, can distort the shape of the event and can either make two cascades appear more like a single cascade or elongate a true single cascade to look more like a double cascade. These optical properties are included in the modeling of the ice, which is included in MC. 
The ice model is based on calibration campaigns from the IceCube experiment~\cite{2020JInst..15P6032A}. 
In addition to this, we include light absorption, scattering, and anisotropy as nuisance parameters in the fit, which accounts for uncertainties in these parameters. For all three of these systematic parameters, the nominal expectations lie within the $1-2\sigma$ regions of the fit. The nominal expectations, allowed range and assumed priors are described in~\cite{BreakPRD}. 
During this work, an ice model which also includes birefringent light propagation in ice~\cite{LEDCalibration}, and we conducted tests with a small sample of simulations with this updated ice model. The tests indicated that the impact of ice systematics is subdominant compared to the large statistical uncertainties for the double cascades sample.
%
%
\nocite{*}
\bibliography{references}

\end{document}

%% file: authorlist.tex
\affiliation{III. Physikalisches Institut, RWTH Aachen University, D-52056 Aachen, Germany}
\affiliation{Department of Physics, University of Adelaide, Adelaide, 5005, Australia}
\affiliation{Dept. of Physics and Astronomy, University of Alaska Anchorage, 3211 Providence Dr., Anchorage, AK 99508, USA}
\affiliation{School of Physics and Center for Relativistic Astrophysics, Georgia Institute of Technology, Atlanta, GA 30332, USA}
\affiliation{Dept. of Physics, Southern University, Baton Rouge, LA 70813, USA}
\affiliation{Dept. of Physics, University of California, Berkeley, CA 94720, USA}
\affiliation{Lawrence Berkeley National Laboratory, Berkeley, CA 94720, USA}
\affiliation{Institut f{\"u}r Physik, Humboldt-Universit{\"a}t zu Berlin, D-12489 Berlin, Germany}
\affiliation{Fakult{\"a}t f{\"u}r Physik {\&} Astronomie, Ruhr-Universit{\"a}t Bochum, D-44780 Bochum, Germany}
\affiliation{Universit{\'e} Libre de Bruxelles, Science Faculty CP230, B-1050 Brussels, Belgium}
\affiliation{Vrije Universiteit Brussel (VUB), Dienst ELEM, B-1050 Brussels, Belgium}
\affiliation{Dept. of Physics, Simon Fraser University, Burnaby, BC V5A 1S6, Canada}
\affiliation{Department of Physics and Laboratory for Particle Physics and Cosmology, Harvard University, Cambridge, MA 02138, USA}
\affiliation{Dept. of Physics, Massachusetts Institute of Technology, Cambridge, MA 02139, USA}
\affiliation{Dept. of Physics and The International Center for Hadron Astrophysics, Chiba University, Chiba 263-8522, Japan}
\affiliation{Department of Physics, Loyola University Chicago, Chicago, IL 60660, USA}
\affiliation{Dept. of Physics and Astronomy, University of Canterbury, Private Bag 4800, Christchurch, New Zealand}
\affiliation{Dept. of Physics, University of Maryland, College Park, MD 20742, USA}
\affiliation{Dept. of Astronomy, Ohio State University, Columbus, OH 43210, USA}
\affiliation{Dept. of Physics and Center for Cosmology and Astro-Particle Physics, Ohio State University, Columbus, OH 43210, USA}
\affiliation{Niels Bohr Institute, University of Copenhagen, DK-2100 Copenhagen, Denmark}
\affiliation{Dept. of Physics, TU Dortmund University, D-44221 Dortmund, Germany}
\affiliation{Dept. of Physics and Astronomy, Michigan State University, East Lansing, MI 48824, USA}
\affiliation{Dept. of Physics, University of Alberta, Edmonton, Alberta, T6G 2E1, Canada}
\affiliation{Erlangen Centre for Astroparticle Physics, Friedrich-Alexander-Universit{\"a}t Erlangen-N{\"u}rnberg, D-91058 Erlangen, Germany}
\affiliation{Physik-department, Technische Universit{\"a}t M{\"u}nchen, D-85748 Garching, Germany}
\affiliation{D{\'e}partement de physique nucl{\'e}aire et corpusculaire, Universit{\'e} de Gen{\`e}ve, CH-1211 Gen{\`e}ve, Switzerland}
\affiliation{Dept. of Physics and Astronomy, University of Gent, B-9000 Gent, Belgium}
\affiliation{Dept. of Physics and Astronomy, University of California, Irvine, CA 92697, USA}
\affiliation{Karlsruhe Institute of Technology, Institute for Astroparticle Physics, D-76021 Karlsruhe, Germany}
\affiliation{Karlsruhe Institute of Technology, Institute of Experimental Particle Physics, D-76021 Karlsruhe, Germany}
\affiliation{Dept. of Physics, Engineering Physics, and Astronomy, Queen's University, Kingston, ON K7L 3N6, Canada}
\affiliation{Department of Physics {\&} Astronomy, University of Nevada, Las Vegas, NV 89154, USA}
\affiliation{Nevada Center for Astrophysics, University of Nevada, Las Vegas, NV 89154, USA}
\affiliation{Dept. of Physics and Astronomy, University of Kansas, Lawrence, KS 66045, USA}
\affiliation{Centre for Cosmology, Particle Physics and Phenomenology - CP3, Universit{\'e} catholique de Louvain, Louvain-la-Neuve, Belgium}
\affiliation{Department of Physics, Mercer University, Macon, GA 31207-0001, USA}
\affiliation{Dept. of Astronomy, University of Wisconsin{\textemdash}Madison, Madison, WI 53706, USA}
\affiliation{Dept. of Physics and Wisconsin IceCube Particle Astrophysics Center, University of Wisconsin{\textemdash}Madison, Madison, WI 53706, USA}
\affiliation{Institute of Physics, University of Mainz, Staudinger Weg 7, D-55099 Mainz, Germany}
\affiliation{Department of Physics, Marquette University, Milwaukee, WI 53201, USA}
\affiliation{Institut f{\"u}r Kernphysik, Universit{\"a}t M{\"u}nster, D-48149 M{\"u}nster, Germany}
\affiliation{Bartol Research Institute and Dept. of Physics and Astronomy, University of Delaware, Newark, DE 19716, USA}
\affiliation{Dept. of Physics, Yale University, New Haven, CT 06520, USA}
\affiliation{Columbia Astrophysics and Nevis Laboratories, Columbia University, New York, NY 10027, USA}
\affiliation{Dept. of Physics, University of Oxford, Parks Road, Oxford OX1 3PU, United Kingdom}
\affiliation{Dipartimento di Fisica e Astronomia Galileo Galilei, Universit{\`a} Degli Studi di Padova, I-35122 Padova PD, Italy}
\affiliation{Dept. of Physics, Drexel University, 3141 Chestnut Street, Philadelphia, PA 19104, USA}
\affiliation{Physics Department, South Dakota School of Mines and Technology, Rapid City, SD 57701, USA}
\affiliation{Dept. of Physics, University of Wisconsin, River Falls, WI 54022, USA}
\affiliation{Dept. of Physics and Astronomy, University of Rochester, Rochester, NY 14627, USA}
\affiliation{Department of Physics and Astronomy, University of Utah, Salt Lake City, UT 84112, USA}
\affiliation{Dept. of Physics, Chung-Ang University, Seoul 06974, Republic of Korea}
\affiliation{Oskar Klein Centre and Dept. of Physics, Stockholm University, SE-10691 Stockholm, Sweden}
\affiliation{Dept. of Physics and Astronomy, Stony Brook University, Stony Brook, NY 11794-3800, USA}
\affiliation{Dept. of Physics, Sungkyunkwan University, Suwon 16419, Republic of Korea}
\affiliation{Institute of Physics, Academia Sinica, Taipei, 11529, Taiwan}
\affiliation{Dept. of Physics and Astronomy, University of Alabama, Tuscaloosa, AL 35487, USA}
\affiliation{Dept. of Astronomy and Astrophysics, Pennsylvania State University, University Park, PA 16802, USA}
\affiliation{Dept. of Physics, Pennsylvania State University, University Park, PA 16802, USA}
\affiliation{Dept. of Physics and Astronomy, Uppsala University, Box 516, SE-75120 Uppsala, Sweden}
\affiliation{Dept. of Physics, University of Wuppertal, D-42119 Wuppertal, Germany}
\affiliation{Deutsches Elektronen-Synchrotron DESY, Platanenallee 6, D-15738 Zeuthen, Germany}

\author{R. Abbasi}
\affiliation{Department of Physics, Loyola University Chicago, Chicago, IL 60660, USA}
\author{M. Ackermann}
\affiliation{Deutsches Elektronen-Synchrotron DESY, Platanenallee 6, D-15738 Zeuthen, Germany}
\author{J. Adams}
\affiliation{Dept. of Physics and Astronomy, University of Canterbury, Private Bag 4800, Christchurch, New Zealand}
\author{S. K. Agarwalla}
\thanks{also at Institute of Physics, Sachivalaya Marg, Sainik School Post, Bhubaneswar 751005, India}
\affiliation{Dept. of Physics and Wisconsin IceCube Particle Astrophysics Center, University of Wisconsin{\textemdash}Madison, Madison, WI 53706, USA}
\author{J. A. Aguilar}
\affiliation{Universit{\'e} Libre de Bruxelles, Science Faculty CP230, B-1050 Brussels, Belgium}
\author{M. Ahlers}
\affiliation{Niels Bohr Institute, University of Copenhagen, DK-2100 Copenhagen, Denmark}
\author{J.M. Alameddine}
\affiliation{Dept. of Physics, TU Dortmund University, D-44221 Dortmund, Germany}
\author{S. Ali}
\affiliation{Dept. of Physics and Astronomy, University of Kansas, Lawrence, KS 66045, USA}
\author{N. M. Amin}
\affiliation{Bartol Research Institute and Dept. of Physics and Astronomy, University of Delaware, Newark, DE 19716, USA}
\author{K. Andeen}
\affiliation{Department of Physics, Marquette University, Milwaukee, WI 53201, USA}
% ...
